\documentclass[aps,twocolumn,showpacs]{revtex4}
\usepackage[dvips]{graphicx}
\usepackage{amsmath}
\usepackage{amsfonts}
\usepackage{amssymb}
\usepackage[linktocpage,breaklinks=true,colorlinks=true,linkcolor=black,filecolor=black,citecolor=black,pagecolor=black,urlcolor=black]{hyperref}

\usepackage{color}

\newcommand{\eq}[1]{\begin{equation}{#1}\end{equation}}

\newcommand{\dd}{\mathrm{d}}
\newcommand{\fKs}{f_{\mathrm{K},s}}
\newcommand{\kB}{k_\mathrm{B}}
\newcommand{\omrec}{\omega_{\mathrm{R},1}}
\newcommand{\Dc}{\Delta_\mathrm{c}}

\begin{document}

\title{Selforganisation and sympathetic cooling of multispecies ensembles in a cavity}
\date{\today}

\author{Tobias Grie{\ss}er}
\affiliation{Institut f\"ur Theoretische Physik, Universit\"at Innsbruck, Technikerstra{\ss}e~25, 6020~Innsbruck, Austria}

\author{Wolfgang Niedenzu}
\affiliation{Institut f\"ur Theoretische Physik, Universit\"at Innsbruck, Technikerstra{\ss}e~25, 6020~Innsbruck, Austria}
 
\author{Helmut Ritsch}
\email{Helmut.Ritsch@uibk.ac.at}
\affiliation{Institut f\"ur Theoretische Physik, Universit\"at Innsbruck, Technikerstra{\ss}e~25, 6020~Innsbruck, Austria}

\begin{abstract}
We predict concurrent selforganisation and cooling of multispecies ensembles of laser-illuminated polarisable particles within a high-$Q$ cavity mode. Resonant collective scattering of laser light into the cavity creates optical potentials which above a threshold pump power transforms a homogeneous particle distribution to a crystalline order for all constituents. Adding extra particles of any mass and temperature always lowers the pump power required for selfordering and allows to concurrently trap atoms, for which high phase-space densities are readily available, in combination with many other kind of atoms, molecules or even polarisable nanoparticles. Collective scattering leads to energy exchange between the different species without direct collisional interactions. We analytically calculate the threshold condition, energy fluxes and the resulting equilibrium phase-space distributions and show that cavity-mediated energy transfer enhances cooling of heavy particles by adding light particles forming a cold reservoir. Extensive numerical many-body simulations support the results of our kinetic analytic model.
\end{abstract}

\pacs{37.30.+i, 37.10.-x, 51.10.+y}

\maketitle
\section{Introduction}
Laser-induced light forces are routinely used to trap and manipulate a large class of polarisable particles from atoms, molecules up to larger objects as microspheres, nanoparticles or even protozoae~\cite{thalhammer2011optical}. Laser cooling, however, has so far been limited to a finite class of atomic species~\cite{metcalf1999laser}, very few kinds of molecules~\cite{doyle2004quo} or individual vibration modes of nanomechanical objects~\cite{kippenberg2008cavity}. Successful cooling requires specific setups with well-chosen laser frequencies and field configurations so that the number of laser-cooled species has only slowly grown in the past years~\cite{shuman2010laser}.
\par
In principle selforganisation and cooling by coherent light scattering in cavities provides a general alternative to trap and cool any kind of polarisable particles, which can be injected into an optical resonator~\cite{vuletic2000laser,lev2008prospects,nimmrichter2010master}. In practise, however, the required particle phase-space densities and laser intensities to reach a useful regime so far have been achieved only for atomic ensembles~\cite{vuletic2001three,slama2007superradiant,baumann2010dicke}, where the theoretical predictions were fully confirmed and showed very fast cooling to sub-Doppler temperatures~\cite{black2003observation}. However, as the required phase-space densities are hard to achieve for molecules~\cite{deachapunya2008slow}, we propose here to generalise the scheme by introducing ensembles of different species and temperatures simultaneously into a single optical resonator. We show that under quite general conditions all species are simultaneously trapped and cooled using only a single laser frequency without the need of direct interparticle interaction. As our central claim we predict that the simultaneous presence of any other species will always increase the total scattering rate for each particle and thus improve the total performance of cooling and trapping for each individual species. As the most interesting case we study a mixture of a precooled and dense enough atomic ensemble with a hotter and much smaller ensemble of molecules or nanospheres. While it would be impossible to reach the selforganisation threshold for the latter alone, combined trapping and sympathetic cooling can be readily achieved without the need of collisions. In fact the different particles could be trapped at different locations within the cavity. Overall this provides for a general route to cool new particle species and also allows a simple general setup for simultaneous multispecies cooling and trapping without the need of a tailored laser for each species. As for a single species, multispecies cooling can be significantly improved using several cavity modes~\cite{domokos2002dissipative,lev2008prospects}. Similarly, a combination of traditional laser cooling methods for atoms with sympathetic cavity cooling could be envisaged to enhance the performance of the combined system.
\par
The paper is organised as follows: After introducing the basic setup and model equations, we calculate the required threshold intensity and joint phase-space density to reach selforganisation and enter the superradiant scattering regime. Then we derive approximate expressions for the cooling dynamics and the energy flow between the different ensembles as well as the asymptotic equilibrium phase-space distributions. These results are finally checked by numerical simulations of the selforganisation and cooling dynamics using a particle model.

\begin{figure}
{\includegraphics[width=7cm]{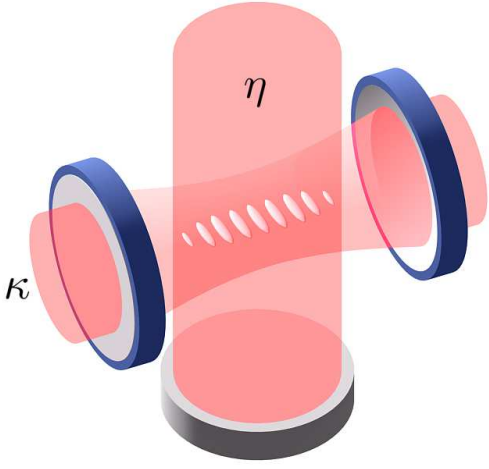}}
\caption{\label{fig1}(Colour online) Setup. A multispecies ensemble of particles within a cavity transversely illuminated by a laser close to resonance with a single cavity mode. Above threshold the particles order in a regular pattern which optimise scattering into the resonator.}
\end{figure}

\section{Model}
Let us consider a mixed dilute gas of $S$ species of $N_s$ ($s=1,\ldots,S)$ polarisable particles each inside a high-$Q$ optical resonator. They are illuminated by a standing wave of counterpropagating off-resonant laser beams that transversely cross the cavity and are close to resonance with a single mode as sketched in fig.~1. Particles within the overlap region of pump laser and cavity mode resonantly scatter light from the one into the other. Pump laser and cavity light give rise to a dynamical optical potential for each species, modifying the particles' distributions. For simplicity we approximate the pump field in the interaction region by a plane standing wave and consider particle motion along the cavity axis only. In single species calculations this approximation proved sufficient to explain the essential physics of selforganisation and cooling~\cite{asboth2005self,vuletic2001three, salzburger2009collective}. An almost ideal implementation of this model could be realised by confining the multiple ensembles into one-dimensional tubes created by two crossed retro-reflected pump laser beams~\cite{haller2009realization}. Extensions of the theory presented in this work to 3d-motion and spatially dependent mode functions are straight-forward and lead to only minor quantitative changes in the appropriate limits.
\par
In terms of the effective laser pump amplitude $\eta_s$, the light shift per photon $U_{0,s}$ and the cavity field amplitude $\alpha$, the combined optical potential for the particles along the cavity axis is given by~\cite{asboth2005self}
\eq{\Phi_s=\hbar \eta_{s}\left(\alpha+\alpha^*\right)\sin(kx)+\hbar U_{0,s}|\alpha|^2\sin^2(kx).}
Combining the one-particle position and momentum variables $z=(x,p)$ and introducing the one-particle Hamiltonian function
\eq{H_s(z,\alpha,\alpha^*)=\frac{p^2}{2m_s}+\Phi_s(x,\alpha,\alpha^*),}
the semi-classical model equations~\cite{domokos2001semiclassical} in Klimontovich's formulation~\cite{klimontovich1995statistical} then read ($s=1,\ldots,S$)
\begin{subequations}\label{Klimontovich}
\eq{\frac{\partial \fKs}{\partial t}+\frac{\partial H_s}{\partial p}\frac{\partial \fKs}{\partial x}-\frac{\partial H_s}{\partial x}\frac{\partial \fKs}{\partial p}=0.}
\eq{\label{eq.field}\dot \alpha=(-\kappa+i\Dc)\alpha-i\sum_{s=1}^S\frac{N_s}{\hbar}\int\frac{\partial H_s}{\partial \alpha^*}\fKs\,\dd^2z-\sqrt{\kappa}~\xi.}
\end{subequations}
Here, $\fKs(z,t)$ is the so-called Klimontovich distribution satisfying $\fKs(z,0)=N_s^{-1}\sum_{j_s=1}^{N_s}\delta(z\!-\!z_{j_s})$ with $\left\{z_{j_s}\right\}$ a set of initial phase points, $\xi$ denotes white noise modelling the fields' quantum fluctuations with $\left\langle \xi(t)\xi^*(t')\right\rangle=\delta(t-t')$, $\left\langle \xi(t)\xi(t')\right\rangle=0$. $\kappa>0$ designates the cavity decay rate and $\Dc=\omega_{\mathrm{p}}-\omega_\mathrm{c}$ is the mismatch between the pump frequency $\omega_\mathrm{p}$ and the cavity resonance frequency $\omega_\mathrm{c}$. The equations~\eqref{Klimontovich} are equivalent to a set of stochastic differential equations (SDEs) for the particles' positions and momenta $z_{j_s}(t)$ and the mode amplitude $\alpha(t)$.

\section{Selforganisation threshold}
Let us decompose the distributions according to ${\fKs(z,t)= f_{s}(z,t)+\delta f_s(z,t)}$ with $f_{s}(z,t)=\left\langle \fKs(z,t)\right\rangle$ denoting the average over a statistical ensemble of similar initial conditions $\left\{z_{j_s}\right\}$, $\alpha(0)$ and the realisations of the noise process $\xi$. The ensemble-averaged Klimontovich distributions, called one-particle distribution functions, fulfil
\eq{\label{mean}\frac{\partial f_s}{\partial t}+\frac{p}{m_s}\frac{\partial f_s}{\partial x}-\frac{\partial\left\langle \Phi_s\right\rangle}{\partial x}\frac{\partial f_s}{\partial p}=\left\langle \frac{\partial \delta \Phi_s}{\partial x}\frac{\partial \delta f_s}{\partial p}\right\rangle.}
These equations are exact but not particularly useful as such because they do not form a closed set. However, for $N_s\rightarrow \infty$, statistical correlations become negligible during the initial stage of the time evolution and hence the one-particle distributions satisfy Vlasov's equation~\cite{vlasov1945kinetic}
\begin{subequations}\label{VlasovKin}
\eq{\label{Vlasov}\frac{\partial f_s}{\partial t}+\frac{p}{m_s}\frac{\partial f_s}{\partial x}-\frac{\partial\Phi_s}{\partial x}\frac{\partial f_s}{\partial p}=0,}
where $\Phi_s=\Phi_s(x,\langle\alpha\rangle,\langle\alpha^*\rangle)$. In the rest of this work we shall for convenience omit the ensemble-average brackets.
These equations together with the average form of eq.~\eqref{eq.field} for the ensemble-averaged mode amplitude
 \begin{multline}\dot \alpha=(-\kappa+i\Dc)\alpha-\\-i\sum_{s}N_s\iint\left[U_{0,s}\alpha\sin^2(kx)+\eta_{s}\sin(kx)\right]f_{s}(x,p)\,\dd x \dd p.
\label{eq.field1}\end{multline}
\end{subequations}represent the essence of the Vlasov kinetic theory of polarisable particles in a resonator describing the initial evolution purely due to the mean field interaction~\cite{griesser2010vlasov}.

\begin{figure}
{\includegraphics[width=8.5cm]{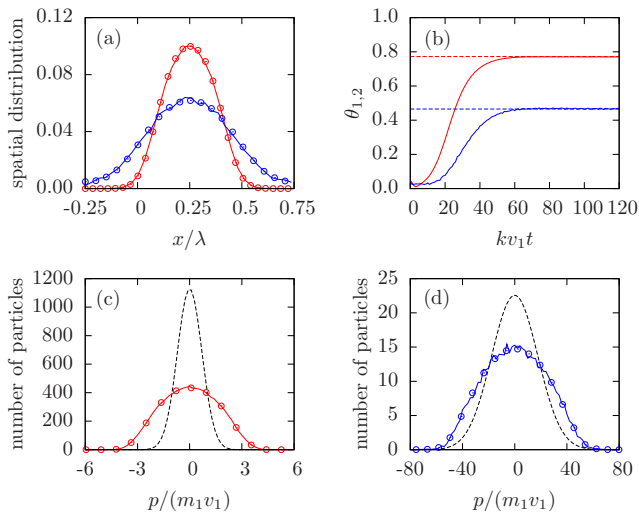}}
\caption{(Colour online) Joint selforganisation of two species starting from a perturbed uniform state above the instability threshold~\eqref{threshold}, such that species one is six times critical, whereas species two is far below its proper threshold. Figure (a) shows the position distributions in the final state, (c) and (d) the momentum distributions initially (dashed) and after selforganisation (solid) as determined from numerical simulations of the particle system~\cite{niedenzu2011kinetic}. The circles show the predictions of eq.~\eqref{adiabatic1}. (b) shows the evolution of the individual order parameters $\theta_s=|\int f_s \sin(kx)\dd x\dd p|$ and the corresponding final values given by the adiabatic theory. Parameters: $N_1=10^4$, $N_2=500$, $m_2=10m_1$, $\kB T_1=10^4\hbar\kappa$, $\kB T_2=2.5\times 10^5\hbar\kappa$, $\eta_1=2.4\kappa$, $\eta_2=27.4\kappa$ and $\omrec=10^{-2}\kappa$.}\label{adiabatic}
\end{figure}

Note that spatially homogeneous initial distributions $f_s(x,p,0)=f_{0,s}(p)$ with zero cavity field $\alpha(0)=0$ are equilibrium states of~\eqref{VlasovKin}. In any finite ensemble, however, density fluctuations cause light scattering and the particles experience friction and diffusion. This is mathematically described by the correlation term in~\eqref{mean}, which induces a slow ``collisional'' evolution of the Vlasov equilibria~\eqref{VlasovKin} towards a new equilibrium. As a central question we now determine the stability of uniform equilibria, i.e.\ whether small perturbations are damped or amplified in the course of time. Linearising equations~\eqref{VlasovKin} around a spatially homogeneous steady state and solving the resulting equations according to Landau~\cite{landau1946vibrations}, one can find the conditions for a dynamical instability under quite general conditions. Obviously, for trapping to occur the effective detuning must be negative, i.e.\ $\delta:=\Dc-\frac{1}{2}\sum_sN_sU_{0,s}<0$.
For convenience we rescale the steady-state distributions as $f_{0,s}(p)=(Lm_sv_{s})^{-1}G_s\big(\frac{p}{mv_{s}}\big)$, where $v_s>0$ is a typical velocity of the $s$th species, $L$ is the cavity length and we assume that these distributions decay monotonously with $|p|$. Then, such an equilibrium is unstable if and only if
\eq{\label{threshold}\sum_{s=1}^S\frac{N_s\eta_{s}^2}{\kB T_s}\left(\mathrm{P}\int_{-\infty}^\infty \frac{G_s'(u)}{-2 u}\dd u\right)>\frac{\kappa^2+\delta^2}{\hbar|\delta|},}
where P denotes the Cauchy principal value and $\kB T_s=m_sv_s^2/2$. In that case, initial density perturbations amplify and the cavity mode amplitude grows exponentially in time at a rate $\gamma>0$ that solves
\eq{(\gamma+\kappa)^2+\delta^2=\sum_{s=1}^S\frac{N_s\eta_s^2\hbar \delta}{2\kB T_s}\int_{-\infty}^\infty\frac{u\, G_{s}'(u)\dd u}{\left(\gamma/kv_s\right)^2+u^2}.\label{eq:gamma1}}
This growth finally ceases and particles and field reach a quasistationary selforganised state. As a central result of this work let us emphasise here, that the right hand side of eq.~\eqref{threshold} only depends on cavity parameters and all terms in the sum on the left hand side are manifestly positive. Hence adding any extra species will always lower the power needed to start the selforganisation process, regardless of temperature and polarisability of the additional particles.
\par
For thermal momentum distributions the integrals in~\eqref{threshold} are unity and the condition gets particularly simple. At higher temperatures, where $(k\min v_s)^2\gg\kappa^2+\delta^2$, the first term in the denominators of eq.~\eqref{eq:gamma1} can be neglected and the field amplitude's growth rate is given by
\eq{\label{gamma}\gamma=-\kappa+\bigg(\sum_s\frac{\hbar|\delta|}{\kB T_s}N_s\eta_s^2-\delta^2\bigg)^{1/2}.}
A glance at this expression shows, that the instability also grows at a larger rate the more terms contribute to the sum. Hence, both the required power and time needed to achieve selforganisation is lowered by combining several species. Obviously, if one can reach the threshold with one species, the system certainly still selforganises if one adds a second species. Let us remark that there exists a dynamical instability for positive effective frequency mismatch $\delta$ as well, but it is connected to heating and does not lead to an ordered distribution. Numerical simulations indicate that the quasi-equilibrium state into which the system evolves in case of instability is close to a BGK solution~\cite{bernstein1957exact} of~\eqref{Vlasov} and~\eqref{eq.field1}. A BGK solution is a stationary solution of~\eqref{VlasovKin} where all one-particle distributions depend on position and momentum solely via the Hamiltonian functions $f_s(z,t)=F_{s}\left(H_s\right)$ and $\alpha(t)=\alpha_0$. The real-valued functions $F_s$ are essentially arbitrary and the steady-state mode amplitude $\alpha_0$ needs to be self-consistently determined from~\eqref{eq.field1}. Let us note that in the weak coupling regime, i.e.\ $\left|\sum_sN_sU_{0,s}\right|\ll|\delta|$, the single-particle actions~\eqref{action} are nearly invariant during selforganisation for a wide range of parameters and in this case it is therefore possible to relate the functions $F_{s}$ to the unperturbed uniform and unstable states $f_{0,s}(p)$ to obtain the selforganised state $f_{s}^{\mathrm{so}}(z)$ as
\eq{\label{adiabatic1}f_{s}^{\mathrm{so}}(x,p)= f_{0,s}\left(J_s\right)}
where $J_s=kI_s$ for untrapped and $J_s=kI_s/2$ for trapped orbits~\cite{krapchev1980adiabatic}. For an illustration the reader may consult fig.~\ref{adiabatic}.

\section{Kinetic theory for the cooling and energy flux}\label{Vlasi}
Let us now turn to describe the system evolution beyond eqs.~\eqref{VlasovKin} in the weak-coupling limit where $\delta\approx\Dc$. To this end we introduce the single-particle action belonging to the instantaneous average potential seen by the $s$th species,
\eq{\label{action}I_s=\pm\frac{1}{2\pi}\oint\sqrt{2m_s\left[H_s-\Phi_s(x')\right]}\,\dd x',}
in which $\Phi_s(x,\alpha)\!\simeq\! 2\hbar\eta_s\mathrm{Re}\left(\alpha\right)\sin(kx)$ as a valid approximation in this limit. The corresponding angle variable $\theta_s$ can be obtained from the generating function $S_s=\pm\int^x \sqrt{2m_s\left[H_s-\Phi_s(x')\right]}\,\dd x'$ as $\theta_s=\frac{\partial S_s}{\partial I_s}$. Starting from any initial condition, in the long time limit the one-particle distribution functions to a good approximation become functions of the single-particle Hamiltonians $H_s$ and thus actions $I_s$ alone. Statistical fluctuations slowly modify these distributions in such a way, that the system evolves towards equilibrium in a sequence of BGK states, $f_s(x,p,t)\simeq f_s(I_s,t)$~\cite{chavanis2007kinetic}. Defining 
\eq{g_{n,s}(I_s,\alpha)=\frac{1}{2\pi}\int_0^{2\pi}\!\sin(kx)e^{-in\theta_s}\,\dd \theta_s}
as well as
\eq{U_s(I_s,\alpha)=\frac{1}{2\pi}\int_0^{2\pi}\frac{\partial I_s}{\partial t}\,\dd \theta_s}
as the average variation of the action along a ``frozen'' orbit, we obtain the following nonlinear Fokker-Planck equations for the one-particle distributions $f_s(I_s,t)$ and the average mode amplitude $\alpha(t)$
\eq{\label{kinetic}\frac{\partial f_s}{\partial t}+U_s\frac{\partial f_s}{\partial I_s}=\frac{\partial }{\partial I_s}\Big(A_s f_s+B_s\frac{\partial f_s}{\partial I_s}+\sum_{s'} C_{BL}[f_s,f_{s'}]\Big)}
\eq{\label{implicit} \alpha=\frac{-2\pi i}{\kappa-i\delta}\sum_s N_s\eta_s\int g_{0,s} f_s~ \dd I_s.}
The collision operator, i.e.\ the r.h.s.\ of~\eqref{kinetic}, consists of two contributions due to the fluctuation and decay of the mode amplitude
\begin{subequations}\label{A}
\eq{A_{s}[f_s]=-4\hbar\delta\eta_s^2\kappa\omega_s\sum_{n}\frac{n^2|g_{n,s}|^2}{|D(in\omega_s)|^2}}
\eq{B_{s}[f_s]=\hbar^2\eta_s^2 \kappa\sum_{n}\frac{n^2|g_{n,s}|^2}{|D(in\omega_s)|^2} \left(\kappa^2+\delta^2+n^2\omega_s^2\right)}
\end{subequations}
and a generalised Balescu-Lenard term~\cite{balescu1960irreversible,lenard1960bogoliubov}
\begin{multline}\label{Balescu-Lenard}C_\mathrm{BL}[f_s,f_l]=8\pi^2\hbar^2\delta^2 N_l\eta_l^2\eta_s^2
 \sum_{n,m}\frac{n|g_{n,s}|^2}{|D(in\omega_s)|^2}\times \\ \times \int|g_{m,l}'|^2\delta\left(n\omega_s-m\omega_l'\right) \Big(n\frac{\partial f_s}{\partial I_s}f_l'-m \frac{\partial f_l}{\partial I_l'}f_s\Big)\dd I_l',
\end{multline}
where the prime means taking the function at $I_l=I_{l}'$ and $\omega_s$ is the nonlinear frequency defined by $\omega_s=\partial H_s/\partial I_s$. For spatially uniform ensembles the actions reduce to $I_s\rightarrow p/k$ and the expressions for the coefficients~\eqref{A} given in~\cite{niedenzu2011kinetic} are recovered. In the above expressions, $D(s)$ denotes the so-called dispersion relation given by
\eq{\label{disp}D(s)=(s+\kappa)^2+\delta^2-i4\pi\hbar\delta\sum_{n,s} N_s\eta_s^2\int\frac{n|g_{n,s}|^2\frac{\partial f_s}{\partial I_s}}{s+in\omega_s}\dd I_s} if $\operatorname{Re}(s)>0$. Apart from the integral terms in the dispersion relation $D$ and the dependence of the nonlinear frequencies $\omega_s$ and functions $g_{n,s}$ and $U_s$ on the common mean field amplitude $\alpha$, the cavity-mediated inter-species interaction is described more explicitly by the Balescu-Lenard collision operator~\eqref{Balescu-Lenard}. It has been derived before by several authors~\cite{luciani1987kinetic,mynick1988generalized} and accounts for the interspecies heat flow, which is mediated by the cavity field. As noted before by these authors $C_\mathrm{BL}$ involves resonant actions $I_l, I_s$, i.e.\ orbits such that the condition $n\omega_l(I_l)=m\omega_s(I_s)$ is fulfilled. This term is a possible source of sympathetic cooling is such a setup.

\subsection{Joint equilibrium states}

\begin{figure}
{\includegraphics[width=6cm]{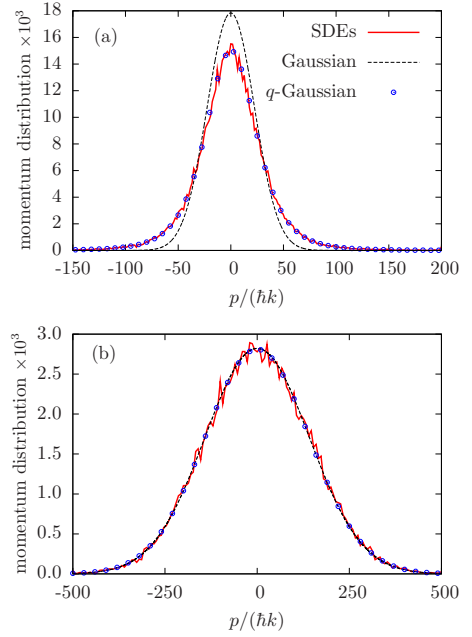}}
\caption{\label{qgauss}(Colour online) Simulated (solid) and analytical (circles) steady-state momentum distributions for two different species $m_2=40 m_1$ below threshold averaged over 250 realisations. The distribution of the lighter particles (a) is given by a $q$-Gaussian with $q_1=1.4$. The dashed curves represent Gaussians corresponding to $\langle p^2\rangle/m_s=\kB T_*$. As $q_2=1.01$ the distribution of the heavier particles (b) is indistinguishable from a Gaussian. Parameters: $N_1=300$, $N_2=200$, $\sqrt{N_1}\eta_1=\sqrt{N_2}\eta_2=800\omrec$, $\kappa=100\omrec$, $\Delta_\mathrm{c}=-2.6\omrec$ and $N_1U_{0,1}=N_2U_{0,2}=-0.1\omrec$.
}
\end{figure}

\begin{figure}
{\includegraphics[width=8cm]{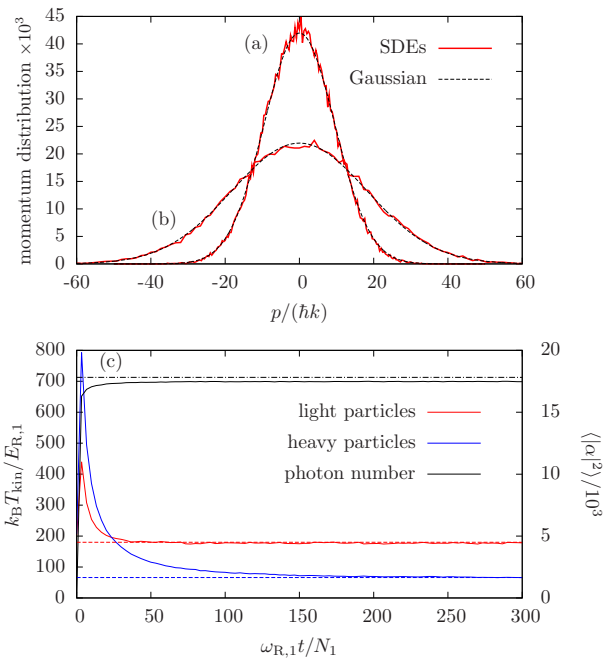}}
\caption{\label{gauss}(Colour online) Selforganised steady-state momentum distributions of (a) species one and (b) species two with $m_2=10m_1$. In (c) we show the time evolution of the kinetic temperatures and the photon number. The dashed lines represent the predictions of eq.~\eqref{EqInhom} and eq.~\eqref{temp1}. The dash-dotted line shows the maximally possible photon number. The initial rise in the kinetic temperatures originates from the fast initial growth of the cavity intensity above threshold. Parameters: $N_1=300$, $N_2=200$, $\sqrt{N_1}\eta_1=\sqrt{N_2}\eta_2=600 \omrec$, $\kappa=100 \omrec$, $\delta=-\kappa$ and $E_{\mathrm{R},1}=\hbar \omrec$.
}
\end{figure}

The set of possible equilibria of~\eqref{kinetic} and~\eqref{implicit} can be divided into two classes: spatially homogeneous with vanishing average field and inhomogeneous with nonzero photon number. The first exist only for effective red detuning $2\delta<-\omega_{\mathrm{R},s}$ and are stable only below the threshold determined by~\eqref{threshold}. The corresponding phase-space distributions can be explicitly calculated to give $q$-Gaussians \eq{\label{Hom}f_{s,\mathrm{eq}}(x,p)\sim \exp_{q_s}\left(-\frac{p^2}{2 m_s\kB T_*}\right),} where the $q$-exponential function is given by
$\mathrm{exp}_q(u)=\left[1+(1-q)u\right]^{\frac{1}{1-q}}$ and we have set
\eq{\label{temp}q_s=1+\frac{\omega_{\mathrm{R},s}}{|\delta|},~~~\kB T_*=\hbar\frac{\kappa^2+\delta^2}{4|\delta|}.}
Their minimal ``thermal'' energy thus is given by $\kB T_*=\hbar \kappa/2$ and is reached if $\delta=-\kappa$. Interestingly, we have a very small but finite minimum kinetic energy in very good resonators but of course if $\kappa\sim\omega_{\mathrm{R},s}$ quantum effects have to be taken into account. As $q_s\rightarrow 1$ the $q$-Gaussian becomes an ordinary Gaussian. For an example the reader is referred to fig.~\ref{qgauss}. It follows that as long as the joint equilibrium $\left\{f_{s,\mathrm{eq}}(x,p), s=1\ldots S\right\}$ is Vlasov-stable according to~\eqref{threshold}, the individual equilibrium distributions are independent of each other because all energy exchange currents cease. The cooling rates, however, are modified by energy exchange between different sorts of particles, described by the generalised Balescu-Lenard operator~\eqref{Balescu-Lenard} and can considerably shorten the time for a given species to reach its steady state. Above threshold, stable and strongly trapped equilibria require $-\delta\gg\omega_{\mathrm{R},s}$ and exist already if the uniform state is only weakly unstable. These are Maxwell-Boltzmann distributions
\eq{ \label{EqInhom} f_{s,\mathrm{eq}}(x,p) \sim \exp \left(-\frac{H_s(z,\alpha,\alpha^*)}{\kB T_{\mathrm{kin},s}}\right),}
with kinetic temperatures
\eq{\label{temp1}\kB T_{\mathrm{kin},s}:=\frac{\langle p^2\rangle}{m_s}=\kB T_*+\hbar\frac{\omega_{0,s}^2}{|\delta|},}
where the trap frequencies are given by $\omega_{0,s}^2=4\eta_s\omega_{\mathrm{R},s}|\operatorname{Re}\,\alpha|$. Again $\alpha$ is determined from~\eqref{implicit} using $g_{0,s}\simeq \frac{k^2 I_s}{2m_s\omega_{0,s}}$ in this limit. Here the equilibria are indeed modified by the presence of additional species through the selfconsistent cavity field but the mutual interaction is not enough to equalise the kinetic temperatures and, as in the uniform case, all equilibrium inter-species heat fluxes vanish. Figure~\ref{gauss} shows an example of a jointly selforganised steady state. Let us note that for any given deeply trapped species the equilibrium uncertainty product $\Delta x\Delta p=\kB T_{\mathrm{kin},s}/\omega_{0,s}$ is bounded from below by $\hbar$,
\eq{\Delta x\Delta p\geq\hbar,}
and thus by twice the minimal value for a particle in a classical potential. The additional uncertainty may therefore be attributed to the quantum fluctuations of the mode amplitude. The energy per particle $E_s$ can be shown to be
\eq{E_s=\Delta x\Delta p \omega_{0,s}\geq \hbar \omega_{0,s}}
which is again twice the usual value. The minimal uncertainty state, which coincides with the minimal energy state, is attained if $2\omega_{0,s}\gg \kappa$ for a detuning $\delta=-2\omega_{0,s}$. These findings remain correct in an entirely quantum-mechanical treatment because for deeply trapped particles and thus approximately harmonic potentials, the semiclassical equations are exactly equivalent to the quantum equations.

\subsection{Sympathetic cooling}
Let us finally examine the energy flow per particle $\dot Q_{2\rightarrow 1}$ from species one to species two for two spatially homogeneous ensembles a little closer. If we assume that species one is already cold, i.e.\ $2\kB T_{\mathrm{kin},1}/\hbar \kappa\ll\kappa/\omrec$ and far from instability, the inter-species heat flow is estimated to be
\begin{multline} \dot Q_{2\rightarrow 1}=-\frac{N_1 m_1}{N_2m_2}\dot Q_{1\rightarrow 2}\simeq N_{1}\eta_2^2\eta_{1}^2\frac{4\sqrt{\pi}\hbar\delta^2}{(\kappa^2+\delta^2)^2} \times \\ \times \sqrt{\frac{\hbar\omrec}{\kB T_{1}}}\left(1-\frac{T_2}{T_{1}}\right)
\left(1+\frac{m_{1}T_2}{m_2T_{1}}\right)^{-3/2},\end{multline}
where we wrote $T_s$ instead of $T_{\mathrm{kin},s}$ for simplicity. Not surprisingly it is maximal if $\delta=-\kappa$. The proportionality of the energy flow to the number of cold particles $N_1$ immediately hints towards a sympathetic cooling scheme, in which a cold ensemble is coupled to a smaller number of hotter and heavier particles whose cooling rate is enhanced due to this exchange current $\dot Q_{2\rightarrow 1}$.
\par
\begin{figure}
{\includegraphics[width=8cm]{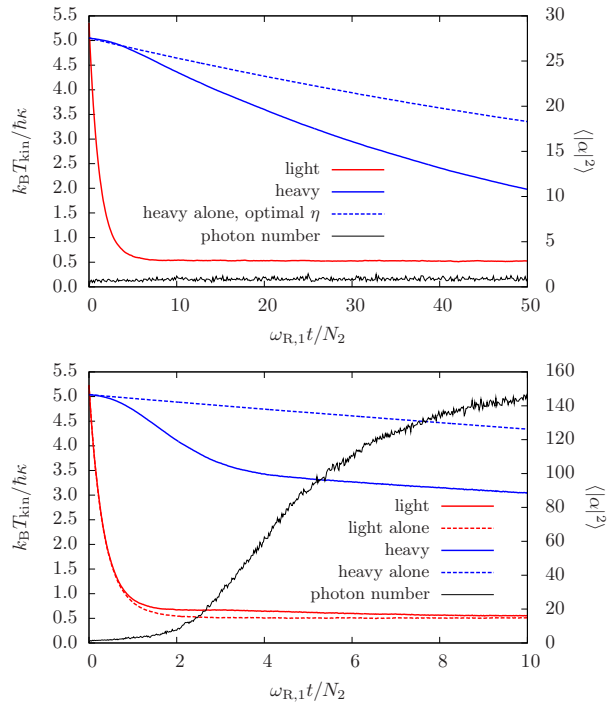}}
\caption{\label{Heat}(Colour online) Time evolution of the kinetic temperatures for two species, one heavy and the other lightweight. Upper plot: Optimal cooling curve of the heavy species alone (blue dashed) vs.\ cooling curve in the presence of a lighter species (blue solid). The cooling is more effective due to the exchange of energy between the species. Parameters: $m_2=200m_1$, $N_1=200$, $N_2=200$, $\sqrt{N_1}\eta_1=134\omrec$, $\sqrt{N_2}\eta_2=134\omrec$, $\kappa=200 \omrec$ and $\delta=-\kappa$. Lower plot: the dashed lines represent the temperature evolutions of each species in the absence of the other species. The sympathetic cooling effect can be observed initially but as the system crosses the selforganisation threshold the inter-species heat flow ceases. This can be inferred from the kinetic temperature curve of the heavier particles (solid blue) becoming parallel to the curve depicting the heavy particles alone (dashed blue). Parameters: $N_1=320$, $N_2=500$, $\sqrt{N_1}\eta_1=207\omrec$, $\sqrt{N_2}\eta_2=258\omrec$, $\kappa=200 \omrec$ and $\delta=-\kappa$.}
\end{figure}

From eq.~\eqref{Balescu-Lenard} one sees that if at least one of the systems is spatially nonuniform (ordered), the inter-species heat flows are effectively suppressed due to the loss of resonances and sympathetic cooling gets inefficient. Hence in order to get a useful inter-species energy transfer one needs to have a large number of cold particles without crossing the instability threshold. This behaviour is exhibited in fig.~\ref{Heat}, where we see that the kinetic energy of the heavy particles decays slower when selforganisation starts and a field is being built up. However, as the field and thus the potential grows the spatial confinement of the particles is continuously increased. This behaviour can be expected to improve using a larger number of non-degenerate modes and several pump frequencies. In this manner, each mode contributes to the energy transfer and dissipation without receiving enough scattered photons to actually trap particles.

\section{Conclusions}
We have shown analytically and in simulations that the selforganisation threshold condition for an ensemble of several different species of particles inside a transversally pumped standing wave resonator is strictly reduced below the value for each of the species separately. Hence the joint selforganisation threshold power is the lower the more species present. If the threshold can be reached with one species, many species can be added and simultaneously trapped and cooled. In the long-time limit the achievable temperatures are only limited by the resonator linewidth and get close to the quantum ground state for deep traps. In the multi-species case the cooling time of a given (heavy) species can be reduced due to energy exchange with a second already colder (light) species. In general one only needs a single intense and narrow laser, frequency-stabilised relative to a cavity mode, to simultaneously trap and cool a large number of different particles within the same volume. The method thus can be applied to atomic gas mixtures, atom-molecule mixtures or even microbeads in a dilute atomic gas. Using more complex setups involving several laser frequencies and modes can be expected to significantly enhance the cooling and lead to more complex distributions of the particles. The method can be easily generalised to moving ensembles in arbitrary cavity geometries, e.g. ring resonators. Here the cavity field mediated interaction of the ensembles transfers a stopping force applied to one ensemble to any other particles. As the general effect has been successfully experimentally demonstrated for single-species setups~\cite{black2003observation,kruse2003cold,ritter2009dynamical}, we are confident that the multispecies generalisation proposed here are well within reach of current technology. This might include even atomic hydrogen, which could be stopped sympathetically with a Helium beam.

\begin{acknowledgements}
We are grateful to Matthias Sonnleitner, Peter Asenbaum and Nikolai Kiesel for stimulating discussions and to Andreas Grie{\ss}er for helping us construct the sketch of the system. We acknowledge support by the Austrian Science Fund FWF through projects SFB FoQuS P13 and P20391.
\end{acknowledgements}

\end{document}